\documentclass[epsfig,showpacs,superscriptaddress]{revtex4}
\usepackage{graphicx}
\usepackage{amssymb}

\begin{document}

\title{A two-level atom coupled to a controllable squeezed vacuum field
reservoir}
\author{Jun-Hong An}
\affiliation{Department of Modern Physics of Lanzhou University, Lanzhou 730000, P. R.
China}
\author{Shun-Jin Wang\footnote{
The corresponding author: gontp@lzu.edu.cn}}
\affiliation{Department of Modern Physics of Lanzhou University, Lanzhou 730000, P. R.
China}
\affiliation{Department of Physics of Sichuan University, Chengdu 610064, P. R. China}
\author{Hong-Gang Luo}
\affiliation{Institute of Theoretical Physics, Chinese Academy of Sciences, Beijing,
100080, P. R. China}
\author{Cheng-Long Jia}
\affiliation{Department of Modern Physics of Lanzhou University, Lanzhou 730000, P. R.
China}

\begin{abstract}
The dissipative and decoherence properties of the two-level atom interacting
with the squeezed vacuum field reservoir are investigated based on the
nonautonomous master equation of the atomic density matrix in the framework
of algebraic dynamics. The nonautonomous master equation is converted into a
Schr\"{o}dinger-like equations and its dynamical symmetry is found based on
the left and right representations of the relevant algebra. The
time-dependent solution and the steady solution are obtained analytically.
The asymptotic behavior of the solution is examined and the approach to the
equilibrium state is proved. Based on the analytic solution the response of
the system to the squeezed vacuum field reservoir is studied numerically.
\end{abstract}

\pacs{03.65.Fd, 03.65.Yz, 42.50.Lc}
\maketitle

\section{Introduction}

The fundamental property of the squeezed state is that the quantum
fluctuations in one quadrature component of the field can be reduced
heavily. Followed the early works \cite{pleb,yue}, much attention
has been given to this specific state. The first experimental result
for the generation of the squeezed state was reported by Slusher
\textit{et al}.\cite{slu} with the scheme of 4-wave mixing in an
optical cavity. Currently a successful scheme for generating
squeezed light can also be based on a parametric oscillator or
parametric down converter \cite{pozik}. Recently, due to its
potential applications in the fields of quantum measurement, optical
communication, and quantum information, the squeezed vacuum state
has been extensively studied \cite{Fur,Hau}. A natural problem is
what effect on physical systems can be induced by the squeezed
vacuum. The squeezed light field will generally be characterized as
a non-stationary reservoir which contains phase dependent features
in the correlation function between pairs of photons. When the
bandwidths of the squeezed lights are not too small, they can be
treated as Markovian reservoirs, and the master equations of the
reduced system can be obtained based on Markovian approximation.

Master equations are of fundamental importance in the treatment of
dissipation and decoherence of open quantum systems. The common feature of
the quantum master equations is the existence of the sandwich terms of the
Liouville operators where the reduced density matrix of the system is in
between some quantum excitation and de-excitation operators. So it is very
difficult to get the exact analytical solution of the master equation, only
simple cases such as a single mode of cavity field coupled to a vacuum state
reservoir ($T=0$) or stationary regime properties are the ones analytically
treated \cite{Scu}. Instead, the master equations are normally converted
into c-number equations in the coherent state representation--the
Fokker-Planck equation \cite{Gar,Wall}. On the other hand, with the
development of the so called quantum engineering, the man-made nonautonomous
quantum systems where the system parameters set by people for controlling of
the system are time-dependent, become more and more important. It would be
very desirable to get the analytical solution of the corresponding master
equation of such a system. Further, even if the total Hamiltonian of an open
system-a system plus an environment, is time-independent, the master
equation of the reduced density matrix of the investigated system still
becomes nonautonomous under the non-Markovian dynamics \cite{Ana}.
Therefore, quantum master equation of the reduced density matrix, in
general, should be nonautonomous.

In the previous works \cite{Zhao,Wang01,An}, we have proposed and developed
an algebraic method to treat the sandwich terms in the Liouville operator
for the nonequilibrium quantum process. This method is just a generalization
of the algebraic dynamical method \cite{Wang93} from quantum mechanical
systems to quantum statistical ones. According to the characteristic of the
sandwich terms in the Liouville operator, the left and right representations
\cite{Wang89} of the relevant algebra have been introduced and the
corresponding composite algebra has been constructed. As a result, the
master equation has been converted into a Schr\"{o}dinger-like equation and
the problems can be solved exactly in the framework of algebraic dynamics.
This method is very effective to treat the nonautonomous quantum system
which contains time-dependent parameters for control of the system.

In this paper, using the algebraic dynamical method, we shall solve the
problem of two-level atom interacting with the squeezed vacuum field
reservoir and investigate what effect on the system can be induced by the
squeezed vacuum field reservoir. As is well known, when the environment is
ordinary thermal equilibrium fields, the model, either be autonomous or be
nonautonomous, has been investigated very well \cite{Gar,An}. When the
environment is the squeezed vacuum field, some work have been done on this
field \cite{phase,Dalt}. Different to Ref. \cite{Dalt}, where two-level
atoms with a time-dependent external classical driven field in the squeezed
vacuum field reservoir were investigated, in this paper, we will concentrate
on the behavior of a two-level atom in a time-dependent squeezed vacuum
field reservoir. With the help of the algebraic dynamical method, the $%
su(2)\oplus su(2)$ dynamical symmetry of the nonautonomous master equation
for the two-level atom is found. The analytical solutions, both the steady
solutions and the time-dependent solutions of the system are obtained
exactly, and the decay property of the atom is investigated. For the
asymptotic behavior of the system, it is proven that any time-dependent
solution of the system approaches its unique steady equilibrium solution.
Based on the analytical solution, the different response behaviors of the
system to the time-dependent squeezing parameter $r$, linear and nonlinear,
are investigated numerically. We noticed that in the nonlinear response
regime, the expectation of one of the squeezed components of the system, $%
\sigma _{y}$, is much enhanced, and the squeezed property of the system,
i.e., the asymmetry of the decay of the expectation of $\sigma _{x}$ and $%
\sigma _{y}$, can also be manifested even when the initial values of $\sigma
_{i}$ ($i=x,y$) are zero.

The paper is organized as follows. In section II, the model Hamiltonian of
the system is presented and the master equation for the reduced matrix of
the atom is deduced. In section III, the dynamical $su(2)_{K}\oplus
su(2)_{J} $ algebraic structure of the Liouville operator of the master
equation is found and the dynamical symmetry of the system is thus exposed.
Section IV is devoted to obtain the analytical non-equilibrium solution of
the nonautonomous master equation, and the approach to the unique steady
equilibrium solution asymptotically is proved. In section V, numerical
results are presented for illustration of non-equilibrium process of some
physical quantities. Discussions and conclusions are given in the final
section.

\section{A two-level atom in the squeezed vacuum field reservoir}

The Hamiltonian of a two-level atom interacting with a squeezed vacuum field
reservoir is
\[
\hat{H}=\frac{1}{2}\hbar \omega _{0}\sigma _{z}+\hbar \sum_{k}\omega
_{k}(a_{k}^{\dagger }a_{k}+\frac{1}{2})+\hbar \sum_{k}g_{k}(\sigma
_{+}a_{k}+h.c.),
\]%
Since the squeezed vacuum field reservoir is kept fixed, the total density
operator of the system can be written as \cite{Scu}%
\[
\rho _{T}(t)=\rho \left( t\right) \otimes \prod_{k}S_{k}\left( \xi \right)
|0_{k}\rangle \langle 0_{k}|S_{k}^{\dagger }\left( \xi \right) ,
\]%
where $S_{k}\left( \xi \right) $ is the squeezing operator and reads as%
\[
S_{k}\left( \xi \right) =\exp (\xi ^{\ast }a_{k_{0}+k}a_{k_{0}-k}-\xi
a_{k_{0}+k}^{\dagger }a_{k_{0}-k}^{\dagger })
\]%
with $\omega =ck_{0}$ and $\xi =r\exp \left( i\theta \right) $, $r$ being
the squeezing parameter and $\theta $ being the reference phase for the
squeezed field. Then the master equation for the reduced density matrix $%
\rho \left( t\right) $\ of the atom interacting with the squeezed vacuum
reservoir can be obtained with the standard Markovian approximation \cite%
{Scu},
\begin{eqnarray}
\dot{\rho} &=&\frac{\gamma }{2}\ (N+1)\ (2\sigma _{-}\rho \sigma _{+}-\sigma
_{+}\sigma _{-}\rho -\rho \sigma _{+}\sigma _{-})  \nonumber \\
&&+\frac{\gamma }{2}\ N\ (2\sigma _{+}\rho \sigma _{-}-\sigma _{-}\sigma
_{+}\rho -\rho \sigma _{-}\sigma _{+})  \nonumber \\
&&-\gamma M\sigma _{-}\rho \sigma _{-}-\gamma M^{\ast }\sigma _{+}\rho
\sigma _{+},  \label{tlmaster}
\end{eqnarray}%
where $\langle a_{k}^{\dagger }a_{k^{\prime }}\rangle =N\delta _{kk^{\prime
}}=\sinh ^{2}(r)\delta _{kk^{\prime }}$, $\langle a_{k}^{\dagger
}a_{k^{\prime }}^{\dagger }\rangle =-M\delta _{k^{\prime },2k_{0}-k}=-\cosh
\left( r\right) \sinh \left( r\right) \exp \left( -i\theta \right) \delta
_{k^{\prime },2k_{0}-k}$, and $N$ is very large. Here we have assumed that
the vacuum field reservoir is the ideal squeezed one, that is, $M^{2}=N(N+1)$%
. When $N\rightarrow \bar{n}=<\frac{1}{\exp (\frac{\hbar \nu _{k}}{k_{B}T})+1%
}><<1$, $M\rightarrow 0$, the master equation (\ref{tlmaster}) reduces to
the familiar form of the master equation \cite{An} describing an two-level
atom coupling to the ordinary thermal equilibrium radiation field. The last
two terms of the above equation exhibit the phase-sensitive nature of the
investigated system. The autonomous case of this equation can be found in
\cite{phase}, while in this paper we concentrate on the nonautonomous case
where all the parameters $M$, $N$ and, $\gamma $ of Eq. (\ref{tlmaster}) are
time-dependent, which allow for adjustment of the squeezing parameter $r$,
the reference phase $\theta $, and the coupling parameter $g_{k}$ during the
time. \

\section{Dynamical symmetry of the master equation}

First we will explore the algebraic structure of Eq. (\ref{tlmaster}). Based
on the left and right representations of certain algebra \cite{Wang01}, we
can get the right and left representations of the algebra $su\left( 2\right)
=\{\sigma _{+},\sigma _{-},\sigma _{z}\}$
\begin{eqnarray}
su\left( 2\right) _{r} &:&[\sigma _{z}^{r},\sigma _{\pm }^{r}]=\pm 2\sigma
_{\pm }^{r},\;\;[\sigma _{+}^{r},\sigma _{-}^{r}]=\sigma _{z}^{r},  \nonumber
\\
su\left( 2\right) _{l} &:&[\sigma _{z}^{l},\sigma _{\pm }^{l}]=\mp 2\sigma
_{\pm }^{l},\;\;[\sigma _{+}^{l},\sigma _{-}^{l}]=-\sigma _{z}^{l}.
\label{basic}
\end{eqnarray}%
It is evident that $su(2)_{r}$ is isomorphic to the $su(2)$, while $%
su(2)_{l} $ anti-isomorphic to the $su(2)$. This is because $su(2)_{r}$
operates towards the right on the bra space $|\ \rangle $; but on the other
hand, $su(2)_{l}$ operates on the ket space $\langle \ |$. Just for the
reason that $su(2)_{r}$ and $su(2)_{l}$ operate on different spaces(the dual
bra and ket spaces), they commute with each other, i.e.%
\begin{equation}
\lbrack su(2)_{r},su(2)_{l}]=0.  \label{com}
\end{equation}%
From these basic representations of $su\left( 2\right) $ we can constitute
two composite algebras%
\begin{eqnarray*}
su(2) &:&\{J_{0}=\frac{\sigma _{z}^{r}+\sigma _{z}^{l}}{2},J_{+}=\sigma
_{+}^{r}\sigma _{-}^{l},J_{-}=\sigma _{-}^{r}\sigma _{+}^{l}\}, \\
su\left( 2\right) &:&\{K_{0}=\frac{\sigma _{z}^{r}-\sigma _{z}^{l}}{2}%
,K_{+}=\sigma _{+}^{r}\sigma _{+}^{l},K_{-}=\sigma _{-}^{r}\sigma _{-}^{l}\}.
\end{eqnarray*}%
According to Eqs. (\ref{basic}) and Eq. (\ref{com}) it's straightforward to
check the following two $su(2)$ commutation relations%
\begin{eqnarray*}
\lbrack J_{0},J_{\pm }] &=&\pm 2J_{\pm },\;\;[J_{+},J_{-}]=J_{0}, \\
\lbrack K_{0},K_{\pm }] &=&\pm 2K_{\pm },\;\;[K_{+},K_{-}]=K_{0}, \\
\lbrack J_{i},K_{j}] &=&0\ \ \ (i,j=0,\pm ).
\end{eqnarray*}%
These two $su(2)$ generators have the action on the bases of von-Neumann
space, which span the basis of the atomic density matrix%
\begin{eqnarray}
J_{0}|s\rangle \langle s^{\prime }| &=&\frac{s+s^{\prime }}{2}|s\rangle
\langle s^{\prime }|,  \nonumber \\
J_{+}|s\rangle \langle s^{\prime }| &=&\delta _{s+1,0}\delta _{s^{\prime
}+1,0}|s+2\rangle \langle s^{\prime }+2|,  \nonumber \\
J_{-}|s\rangle \langle s^{\prime }| &=&\delta _{s-1,0}\delta _{s^{\prime
}-1,0}|s-2\rangle \langle s^{\prime }-2|,  \nonumber \\
K_{0}|s\rangle \langle s^{\prime }| &=&\frac{s-s^{\prime }}{2}|s\rangle
\langle s^{\prime }|,  \nonumber \\
K_{+}|s\rangle \langle s^{\prime }| &=&\delta _{s+1,0}\delta _{s^{\prime
}-1,0}|s+2\rangle \langle s^{\prime }-2|,  \nonumber \\
K_{-}|s\rangle \langle s^{\prime }| &=&\delta _{s-1,0}\delta _{s^{\prime
}+1,0}|s-2\rangle \langle s^{\prime }+2|,  \label{action}
\end{eqnarray}%
where $s(s^{^{\prime }})=\pm 1$.

By virtue of the above algebras, the nonautonomous master equation (\ref%
{tlmaster}) can be rewritten as a linear combination of these generators%
\begin{equation}
\dot{\rho}=\gamma \left( t\right) \{[N(t)+1]J_{-}+N(t)J_{+}-\frac{1}{2}%
J_{0}-M(t)K_{-}-M^{\ast }(t)K_{+}-\frac{2N\left( t\right) +1}{2}\}\rho
=\Gamma \rho .  \label{tlm}
\end{equation}%
which implies that Eq. (\ref{tlmaster}) possesses an $su\left( 2\right)
_{J}\oplus su\left( 2\right) _{K}$ dynamical symmetry. Thus it is integrable
and can be solved analytically according to algebraic dynamics \cite{Wang93}%
. Moreover, the master equation (\ref{tlmaster}) is converted into a Schr%
\"{o}dinger-like equation (\ref{tlm}), where the rate operator $\Gamma $
plays the role of the Hamiltonian and the reduced matrix plays the role of
the wavefunction.

It is noted that Eq. (\ref{tlm}) is a time-dependent generalization of Eq. (%
\ref{tlmaster}) and it is still under the\ Markovian approximation.
That implies a basic assumption: the time dependence of the
parameters of the master equation don't alter the base of the
Markovian approximation-the weak coupling assumption of the system
and the reservoir.

\section{Exact solution to the master equation in the nonautonomous case}

\subsection{Steady solution of the master equation}

To better understand the time-dependent solution of the master equation and
its decay behavior, we first consider the long-time case $\left( \gamma
\left( t\right) \rightarrow \gamma ,N(t)\rightarrow N,M(t)\rightarrow
M\right) $ and the eigen equation problem of the master equation (\ref{tlm}%
). From Eq. (\ref{tlm}) we can get%
\begin{equation}
\ \Gamma \rho =\beta \rho  \label{eigen}
\end{equation}%
Introducing the similarity transformation%
\[
U=e^{\alpha _{+}J_{+}}e^{\alpha _{-}J_{-}}e^{\eta _{+}K_{+}}e^{\eta
_{-}K_{-}},
\]%
and under the transformation parameter conditions%
\begin{eqnarray}
(N+1)\alpha _{+}^{2}+\alpha _{+}-N &=&0,  \nonumber \\
(N+1)(1+2\alpha _{+}\alpha _{-})+\alpha _{-} &=&0,  \label{con} \\
M\eta _{+}^{2}-M^{\ast } &=&0,  \nonumber \\
\ 1+2\eta _{+}\eta _{-} &=&0,  \label{conds}
\end{eqnarray}%
we can transform Eq. (\ref{eigen}) to the diagonal form as
\begin{eqnarray}
\bar{\Gamma}\bar{\rho} &=&\beta \bar{\rho},  \nonumber \\
\bar{\rho} &=&U^{-1}\rho ,  \nonumber \\
\bar{\Gamma} &=&U^{-1}\Gamma U=-\gamma \{[(N+1)\alpha _{+}+\frac{1}{2}%
]J_{0}-M\eta _{+}K_{0}+\frac{2N+1}{2}\}.  \label{ratebar}
\end{eqnarray}%
\ The eigensolutions of \ Eq. (\ref{eigen}) are
\begin{eqnarray}
\beta (s,s^{\prime }) &=&-\gamma \{[(N+1)\alpha _{+}+\frac{1}{2}]\frac{%
s+s^{^{\prime }}}{2}-M\eta _{+}\frac{s-s^{^{\prime }}}{2}+\frac{2N+1}{2}\},
\nonumber \\
\rho (s,s^{\prime }) &=&e^{\alpha _{+}J_{+}}e^{\alpha _{-}J_{-}}e^{\eta
_{+}K_{+}}e^{\eta _{-}K_{-}}|s\rangle \langle s^{\prime }|.
\label{eigensolution}
\end{eqnarray}%
It is interesting to note that Eqs. (\ref{con}) and Eqs. (\ref{conds}) both
have two sets of solutions. The two combinations of the solutions $(\alpha
_{+}=-1,\ \alpha _{-}=(N+1)/(2N+1),\ \eta _{+}=\pm e^{i\theta },\ $and $\eta
_{-}=\mp e^{-i\theta })$ and $(\alpha _{+}=N/(2N+1),\ \alpha
_{-}=-(N+1)/(2N+1),\ \eta _{+}=\pm e^{i\theta },\ $and $\eta _{-}=\mp
e^{-i\theta })$ both contain the same zero-mode solution (the unique steady
equilibrium solution) and they are equivalent physical solutions since all
their eigen values $\beta (s,s^{\prime })$ are non-positive (this can be
proved by the series expansion $M=\sqrt{N(N+1)}\approx N(1+\frac{1}{2N}-\frac{1%
}{8N^{2}})$ and by the fact $N\pm M+\frac{1}{2}>0$).

Noticing that $J_{\pm }$\ yield non-zero results only if they act on
diagonal elements and $K_{\pm }$ do so if they act on non-diagonal
elements, we see that the zero
mode steady solution of the system has the same form (but with different $N$%
) as that in Ref. \cite{An} where the reservoir is a thermal equilibrium
field,%
\begin{equation}
\rho _{0}=\frac{N+1}{2N+1}|-1\rangle \langle -1|+\frac{N}{2N+1}|+1\rangle
\langle +1|,  \label{stead2}
\end{equation}

It is noted that the rate operator $\Gamma $ is non-Hermitian, i.e. $\Gamma
^{\dagger }\neq \Gamma $, which is evident from $J_{+}^{\dagger }=J_{-}$, $%
J_{-}^{\dagger }=J_{+}$, $J_{0}^{\dagger }=J_{0}$, $K_{+}^{\dagger }=K_{-}$,
and $K_{-}^{\dagger }=K_{+}$. Just because of this non-Hermiticity, the
eigenvectors of $\Gamma $ and $\Gamma ^{\dagger }$ constitute a
bi-orthogonal basis \cite{Feshbach}. By introducing a similarity
transformation $U^{\prime }=(U^{-1})^{\dagger }$ and umder the same
conditions as Eqs. (\ref{con}, \ref{conds}), the operator $\Gamma ^{\dagger
} $ can be diagonalized. Then the eigensolutions of $\Gamma ^{\dagger }$ are
given by
\begin{eqnarray}
\Gamma ^{\dagger }\tilde{\rho}(s,s^{\prime }) &=&\tilde{\beta}(s,s^{\prime })%
\tilde{\rho}(s,s^{\prime })  \nonumber \\
\tilde{\beta}(s,s^{\prime }) &=&\beta ^{\ast }(s,s^{\prime })  \nonumber \\
\tilde{\rho}(s,s^{\prime }) &=&e^{-\eta _{+}K_{-}}e^{-\eta
_{-}K_{+}}e^{-\alpha _{+}J_{-}}e^{-\alpha _{-}J_{+}}|s\rangle \langle
s^{\prime }|.  \label{eigensolu}
\end{eqnarray}%
One can easily to check that the eigenvectors $\rho (s,s^{\prime })$ and $%
\tilde{\rho}(s,s^{\prime })$ form a biorthogonal set.~A similar discussion
can be found in Ref. \cite{An}.

To compare with the work of Ref. \cite{Briegel,Sten,Jakob}, we consider the
left eigensolutions of $\Gamma $. The transformed form of the eigen equation
is
\[
\bar{\rho}^{\prime }\bar{\Gamma}=\beta ^{\prime }(s,s^{\prime })\bar{\rho}%
^{\prime }
\]%
Eq.~(\ref{ratebar}) shows that the operator $\bar{\Gamma}=U^{-1}\Gamma U$ is
self-adjoint and, thus, $\bar{\rho}^{\prime }\bar{\Gamma}=U^{\dagger }\Gamma
^{\dagger }(U^{-1})^{\dagger }\bar{\rho}^{\prime }=[(U^{-1})^{\dagger
}]^{-1}\Gamma ^{\dagger }(U^{-1})^{\dagger }\bar{\rho}^{\prime }$. This
implies that $\Gamma ^{\dagger }$ has the same eigensolutions as $\Gamma $
when they act from the left and right on $\rho $, respectively. In the
phrase of Ref. \cite{Briegel}, $\rho (s,s^{\prime })$ is the right-hand
eigenvectors of the operator $\Gamma $ and $\tilde{\rho}(s,s^{\prime })$ is
the right-hand eigenvectors of the operator $\Gamma ^{\dagger }$.
Furthermore the right-hand eigenvectors of the operator $\Gamma ^{\dagger }$
coincide with the left-hand eigenvectors of $\Gamma $. And the two sets of
eigenvectors constitute a biorthogonal set. From this biorthogonal set one
can construct a generalized entropy functional varying monotonically with
time, which named as the Lyapunov functional and have a crucial rules in
control theory \cite{Jakob}.

\subsection{Time-dependent solution of the master equation}

Now we turn back to the nonautonomous case and study the time-dependent
solution of Eq. (\ref{tlm}). With the gauge transformation
\[
U_{g}\left( t\right) =e^{\alpha _{+}\left( t\right) J_{+}}e^{\alpha
_{-}\left( t\right) J_{-}}e^{\eta _{+}\left( t\right) K_{+}}e^{\eta
_{-}\left( t\right) K-},
\]%
and the gauge parameters satisfying
\begin{eqnarray}
\frac{d\alpha _{+}(t)}{dt} &=&-\gamma (t)[N(t)+1]\alpha _{+}^{2}(t)-\gamma
(t)\alpha _{+}(t)+\gamma (t)N(t),  \nonumber \\
\frac{d\alpha _{-}(t)}{dt} &=&\gamma (t)[N(t)+1][1+2\alpha _{+}(t)\alpha
_{-}(t)]+\gamma (t)\alpha _{-}(t),  \nonumber \\
\frac{d\eta _{+}\left( t\right) }{dt} &=&\gamma (t)[M(t)\eta _{+}^{2}\left(
t\right) -M^{\ast }(t)],  \nonumber \\
\frac{d\eta _{-}\left( t\right) }{dt} &=&-\gamma (t)M(t)[1+2\eta _{+}\left(
t\right) \eta _{-}\left( t\right) ].  \label{differ}
\end{eqnarray}%
where the initial conditions of the gauge transformation is $U_{g}(0)=1$,
the operator $\Gamma $ can be transformed to the diagonal form
\begin{eqnarray*}
\bar{\Gamma}\left( t\right) &=&U_{g}^{-1}\left( t\right) \Gamma U_{g}\left(
t\right) -U_{g}^{-1}\left( t\right) \dot{U}_{g}\left( t\right) \\
&=&-\gamma (t)\{[(N\left( t\right) +1)\alpha _{+}\left( t\right) +\frac{1}{2}%
]J_{0}-M\left( t\right) \eta _{+}\left( t\right) K_{0}+\frac{2N\left(
t\right) +1}{2}\}
\end{eqnarray*}%
If the initial state of the system is $\rho \left( 0\right)
=\sum_{s,s^{\prime }}\lambda _{s,s^{\prime }}|s\rangle \langle s^{\prime }|$%
, then the time-dependent solution of the nonautonomous master Eq. (\ref{tlm}%
) is%
\begin{eqnarray}
\rho \left( t\right) &=&\sum_{s,s^{\prime }}\lambda _{s,s^{\prime
}}U_{g}\left( t\right) e^{\int_{0}^{t}-\gamma (\tau )\{[(N\left( \tau
\right) +1)\alpha _{+}\left( \tau \right) +\frac{1}{2}]\frac{s+s^{\prime }}{2%
}-M\left( \tau \right) \eta _{+}\left( \tau \right) \frac{s-s^{\prime }}{2}+%
\frac{2N\left( \tau \right) +1}{2}\}d\tau }|s\rangle \langle s^{\prime }|
\nonumber \\
&=&\{\lambda _{1,1}f_{1,1}(t)[1+\alpha _{+}\left( t\right) \alpha
_{-}(t)]+\lambda _{-1,-1}f_{-1,-1}(t)\alpha _{+}\left( t\right) \}|+1\rangle
\langle +1|  \nonumber \\
&&+[\lambda _{1,1}f_{1,1}(t)\alpha _{-}(t)+\lambda
_{-1,-1}f_{-1,-1}(t)]|-1\rangle \langle -1|  \nonumber \\
&&+\{\lambda _{1,-1}f_{1,-1}(t)[1+\eta _{+}\left( t\right) \eta
_{-}(t)]+\lambda _{-1,1}f_{-1,1}(t)\eta _{+}\left( t\right)
\}|+1\rangle\langle -1|  \nonumber \\
&&+[\lambda _{1,-1}f_{1,-1}(t)\eta _{-}(t)+\lambda
_{-1,1}f_{-1,1}(t)]|-1\rangle \langle +1|,  \label{solut}
\end{eqnarray}%
where we have defined $f_{s,s^{\prime }}(t)=e^{\int_{0}^{t}-\gamma (\tau
)\{[(N\left( \tau \right) +1)\alpha _{+}\left( \tau \right) +\frac{1}{2}]%
\frac{s+s^{\prime }}{2}-M\left( \tau \right) \eta _{+}\left( \tau \right)
\frac{s-s^{\prime }}{2}+\frac{2N\left( \tau \right) +1}{2}\}d\tau }$.

One can check that the solution of Eq. (\ref{solut}) recovers the results of
the autonomous system with time-independent parameters. For the autonomous
case, the explicit solutions of the gauge parameters $\alpha _{\pm }$ and $%
\eta _{\pm }$ can be obtained from Eqs. (\ref{differ})%
\begin{eqnarray*}
\alpha _{+}(t) &=&\frac{1-e^{-\gamma (2N+1)t}}{\frac{N+1}{N}+e^{-\gamma
(2N+1)t}}, \\
\alpha _{-}(t) &=&\frac{(N+1)N[\frac{N+1}{N}+e^{-\gamma
(2N+1)t}][1-e^{-\gamma (2N+1)t}]}{(2N+1)^{2}e^{-\gamma (2N+1)t}}. \\
\eta _{+}(t) &=&\frac{1-e^{2\gamma Mt}}{1+e^{2\gamma Mt}}, \\
\eta _{-}(t) &=&\frac{(1-e^{2\gamma Mt})(1+e^{2\gamma Mt})}{4e^{2\gamma Mt}},
\end{eqnarray*}%
where $M=M^{\ast }$ is assumed. Then under the initial condition of the
system
\begin{equation}
\rho \left( 0\right) =|\mu |^{2}|1\rangle \langle 1|+|\nu |^{2}|-1\rangle
\langle -1|+\mu \nu ^{\ast }|1\rangle \langle -1|+\mu ^{\ast }\nu |-1\rangle
\langle 1|,  \label{inia}
\end{equation}%
the expectation values of $\sigma _{x}$, $\sigma _{y}$, and $\sigma _{z}$ are%
\begin{eqnarray}
\langle \sigma _{x}\rangle &=&(\mu \nu ^{\ast }+\mu ^{\ast }\nu )e^{-\gamma
(N+M+\frac{1}{2})t},  \nonumber \\
\langle \sigma _{y}\rangle &=&\frac{1}{i}(\mu ^{\ast }\nu -\mu \nu ^{\ast
})e^{-\gamma (N-M+\frac{1}{2})t},  \nonumber \\
\langle \sigma _{z}\rangle &=&\frac{2[|\mu |^{2}(N+1)-|\nu |^{2}N]e^{-\gamma
(2N+1)t}-1}{2N+1}.  \label{expecta}
\end{eqnarray}%
So the decoherent characteristic times of $\langle \sigma _{x}\rangle $, $%
\langle \sigma _{y}\rangle $, and $\langle \sigma _{z}\rangle $ are $\frac{1%
}{\gamma (N+M+\frac{1}{2})},$ $\frac{1}{\gamma (N-M+\frac{1}{2})},$and $%
\frac{1}{\gamma (2N+1)}$, respectively, which is same as the results of Ref.
\cite{phase}. The decoherent characteristic time of $\langle \sigma
_{y}\rangle $ is much larger than that of $\langle \sigma _{x}\rangle $.
From the above results we see that the radiative properties of the atom
depend sensitively on the state of the environment to which it is coupled.
The squeezed vacuum reservoir imposes its phase information on the atom and
makes a highly asymmetric decay property of the atom. This is quite
different from thermal vacuum reservoir which results in the same decay rate
(where $M=0$, $N\rightarrow \bar{n}$) for $\langle \sigma _{y}\rangle $ and $%
\langle \sigma _{x}\rangle $.

So we can see that the time-dependent solution of Eq. (\ref{solut})
reproduces all the results of the autonomous case. In the following we shall
concentrate on the nonautonomous case and investigate the time evolution
behaviors of the system.

\subsection{\protect\bigskip The asymptotical behavior of the solution}

In this section we shall examine the asymptotic behavior of the
time-dependent analytical solution (\ref{solut}) of the nonautonomous master
equation and prove that it approaches to the equilibrium steady solution (%
\ref{stead2}). To this end we shall first investigate the asymptotic
behavior of the gauge parameters which are solutions of Eqs. (\ref{differ})
from which one notices that $d\eta _{+}(t)/dt<0(>0)$ if $-1<\eta _{+}(t)<0$
(if $\eta _{+}(t)<-1$ and $\eta _{+}(t)>0$). With the initial condition $%
\eta _{+}\left( 0\right) =0$, we see that $\eta _{+}\left( t\right) $
approaches the value $\eta _{+}\left( \infty \right) =-1$ asymptotically. To
get the asymptotic solution of $\eta _{-}\left( t\right) $, we define $%
y(t)=\eta _{-}\left( t\right) e^{\int_{0}^{t}2\gamma (\tau )M\left( \tau
\right) \eta _{+}\left( \tau \right) d\tau }=\eta _{-}\left( t\right)
e^{\int_{0}^{t}p\left( \tau \right) d\tau }$. The time differential of $%
y\left( t\right) $ is given by $b\exp \int_{0}^{t}p(\tau )d\tau $ where $b=%
\dot{\eta}_{-}(t)+\eta _{-}(t)p(t)$. Since $b\longrightarrow \gamma $ which
is bounded and $p(t)$ is negative for large $t$, the differential $\dot{y}%
\left( t\right) $\ tends to zero, hence $y\left( t\right) $ tends
towards a constant. With the same procedure we can get the
asymptotic solutions of $\alpha _{\pm }\left( t\right) $.
Summarizing the above discussion, we have
\begin{eqnarray*}
\alpha _{+}(t)|_{t\rightarrow \infty } &=&\frac{N}{N+1} \\
\alpha _{-}(t)\times e^{-\int_{0}^{t}\gamma (\tau )[N(\tau )+1][\alpha
_{+}(\tau )+1]d\tau }|_{t\rightarrow \infty } &=&const \\
\eta _{+}\left( t\right) |_{t\rightarrow \infty } &=&-1 \\
\eta _{-}\left( t\right) e^{\int_{0}^{t}2\gamma (\tau )M\left( \tau \right)
\eta _{+}\left( \tau \right) d\tau }|_{t\rightarrow \infty } &=&const.
\end{eqnarray*}%
Although, for the steady case, two sets of the parameter solutions $\alpha
_{\pm }$ and $\eta _{\pm }$ \ lead to two equivalent transformations which
diagonalize the rate operator and generate the same physical solution, as
discussed above; for the time-dependent case, the properties of the two sets
of steady solutions of $\alpha _{\pm }$ and $\eta _{\pm }$ are not on equal
footing. It is found that only one set of the solutions can be reached by
the time-dependent solution asymptotically. Using these asymptotic relations
we have the asymptotic results of the time-dependent solution as follows%
\begin{eqnarray*}
\rho _{++}(t)|_{t\rightarrow \infty } &=&e^{-\int_{0}^{t}\gamma (\tau
)[N(\tau )+1][\alpha _{+}(\tau )+1]d\tau }[|+1\rangle \langle +1|+\alpha
_{-}(t)(|-1\rangle \langle -1|+\alpha _{+}(t)|+1\rangle \langle +1|)] \\
&\longrightarrow &const.\times \rho _{0} \\
\rho _{--}(t)|_{t\rightarrow \infty } &=&e^{-\int_{0}^{t}\gamma (\tau )[-(%
\bar{n}_{0}(\tau )+1)\alpha _{+}(\tau )+\bar{n}_{0}(\tau )]d\tau
}[|-1\rangle \langle -1|+\alpha _{+}(t)|+1\rangle \langle
+1|]\longrightarrow const.\times \rho _{0} \\
\rho _{+-}(t)|_{t\rightarrow \infty } &=&e^{-\int_{0}^{t}\gamma (\tau
)[M\left( \tau \right) \eta _{+}\left( \tau \right) +N\left( \tau \right) +%
\frac{1}{2}]d\tau }e^{\int_{0}^{t}2\gamma (\tau )M\left( \tau \right) \eta
_{+}\left( \tau \right) d\tau }\{|+1\rangle \langle -1|+\eta _{-}\left(
t\right) [\eta _{+}\left( t\right) |+1\rangle \langle -1|+|-1\rangle \langle
+1|]\} \\
&\longrightarrow &e^{-\int_{0}^{t}\gamma (\tau )[M\left( \tau \right)
+N\left( \tau \right) +\frac{1}{2}]d\tau }|+1\rangle \langle
-1|+e^{-\int_{0}^{t}\gamma (\tau )[N\left( \tau \right) -M\left( \tau
\right) +\frac{1}{2}]d\tau }\times const[|-1\rangle \langle +1|-|+1\rangle
\langle -1|] \\
&\longrightarrow &0 \\
\rho _{-+}(t)|_{t\rightarrow \infty } &=&e^{-\int_{0}^{t}\gamma (\tau
)[M\left( \tau \right) \eta _{+}\left( \tau \right) +N\left( \tau \right) +%
\frac{1}{2}]d\tau }[\eta _{+}\left( t\right) |+1\rangle \langle
-1|+|-1\rangle \langle +1|] \\
&\longrightarrow &0
\end{eqnarray*}

The above results indicate that any time-dependent solution of the master
equation in the nonautonomous case asymptotically approaches the unique
steady equilibrium solution, irrespective of their initial conditions.

\section{Numerical results}

To study the squeezed properties of the system in the nonautonomous case, we
shall start from the time-dependent analytical solution Eq. (\ref{solut})
and calculate the expectation values of the two-level atomic operators. The
expectation values of $\sigma _{i}\left( i=x,y,z\right) $ are%
\begin{eqnarray}
\langle \sigma _{x}\rangle &=&\frac{\mu \nu ^{\ast }f_{1,-1}[1+\eta
_{+}(t)\eta _{-}(t)+\eta _{-}(t)]+\mu ^{\ast }\nu f_{-1,1}[1+\eta _{+}(t)]}{2%
},  \nonumber \\
\langle \sigma _{y}\rangle &=&\frac{\mu \nu ^{\ast }f_{1,-1}[1+\eta
_{+}(t)\eta _{-}(t)-\eta _{-}(t)]+\mu ^{\ast }\nu f_{-1,1}[\eta _{+}(t)-1]}{%
2i},  \nonumber \\
\langle \sigma _{z}\rangle &=&|\mu |^{2}f_{1,1}[1+\alpha _{+}(t)\alpha
_{-}\left( t\right) -\alpha _{-}\left( t\right) ]+|\nu |^{2}f_{-1,-1}[\alpha
_{+}\left( t\right) -1],  \label{nonexpe}
\end{eqnarray}%
where the initial state is same as Eq. (\ref{inia}). Given any
time-dependent parameters $\gamma (t),\ r(t)$, and $\theta (t)$, we
can get the gauge transformation parameters $\alpha _{\pm }(t)$ and
$\eta _{\pm }$ from Eqs. (\ref{differ}) numerically. Then substitute
$\alpha _{\pm }(t)$ and $\eta _{\pm }$ to Eqs. (\ref{nonexpe}), we
can obtain the time-dependent behaviors of the expectation values
varying according to the system parameters. In the following we will
study the variation of the time-dependent behavior of $ \langle
\sigma _{i}\rangle $ with the squeezing parameter $r(t)$.

Without loss generalization, to explore the time-dependent response of the
system to the squeezing parameters we choose time-dependent parameters as $%
r=c_{1}e^{-c_{2}t}$, $\gamma =1$, and $\theta =0$. Fig. 1 and Fig. 2 show
the time-dependent behaviors of $\langle \sigma _{i}\rangle $ with $%
c_{1}=0.1 $ and $c_{2}=0.1$ under the initial condition $\mu =\sqrt{0.2}%
e^{i/3}$, $\nu =\sqrt{0.8}e^{2\pi i}$ and $\mu =\sqrt{0.2}$, $\nu =\sqrt{0.8}
$ respectively. Fig. 3 and Fig. 4 correspond to $c_{2}=0.3$. The behavior of
$\langle \sigma _{i}\rangle $ varying to $r(t)$ with parameter $c_{2}=0.6$
were plotted on Fig. 5 and Fig. 6. From these figures we see that $\langle
\sigma _{z}\rangle $ decays to its steady value $-1$; $\langle \sigma
_{x}\rangle $ and $\langle \sigma _{y}\rangle $ decay to their steady value $%
0$, but the decay time of $\langle \sigma _{y}\rangle $ is much larger than
that of $\langle \sigma _{x}\rangle $, which regenerate the results of the
autonomous case. From Fig. 2 we notice that $\langle \sigma _{y}\rangle =0$,
so the squeezed property of the squeezed vacuum field to the two-level atom
is covered by the initial value of $\langle \sigma _{y}\rangle $. For the
small squeezing parameter, the time-dependent response of the system with
respect to the squeezing parameter is nearly linear and much like the
autonomous case. However, with the increase of squeezing parameter, the
response is no longer linear and the decay time of $\langle \sigma
_{x}\rangle $'s becomes shorter and shorter, while that of $\langle \sigma
_{y}\rangle $'s becomes longer and longer. A more interesting phenomenon
appearing during the increase of the squeezing parameter is that the value
of $\langle \sigma _{y}\rangle $ can be much enhanced. And the resulting
effect of the squeezed vacuum field reservoir on the two-level atom, i.e.,
the large asymmetric behaviors of decay between $\langle \sigma _{x}\rangle $
and $\langle \sigma _{y}\rangle $, can also be presented even when some of
them are zero initially, which is not presented in the autonomous case (this
can also be seen from the second expression of Eqs. (\ref{expecta})).
Moreover, the time-dependent behavior of $\langle \sigma _{y}\rangle $ is
completely different from that of $r(t)$, and the response of system to the
squeezing parameter thus exhibits nonlinearity.

From the analysis above we see that the squeezed vacuum field
reservoir acts as a squeezing mold and imprint its squeezing
information to the two-level atom, which makes the asymmetry of
the two decay behaviors. With a time-dependent squeezing parameter
this action can be more enhanced, which widen the results of Ref.
\cite{phase} and provide a potential way to get more stable atomic
variable. The similar time-dependent system can also be found in
Ref. \cite{Berg}, where a technique of stimulated Raman adiabatic
passage was used to get a complete population transfer between two
atomic states. There the two adiabatic varying lasers act as a
role of coherent trapping in the three-level atomic system and the
technique relies on the coherent trapping state.

\section{Summary and Outlook}

We have investigated the nonautonomous master equation of the two-level atom
interacting with the squeezed vacuum reservoir for the first time. The new
results of our paper compared to the previous ones are: (1) the dynamical
symmetry of the nonautonomous master equation has been established and the
master equation has been converted into a Schr\"{o}dinger-like equation.
Based on the dynamical symmetry we have obtained the most general
non-equilibrium solutions of the nonautonomous master equation analytically;
and as a special case, our solutions recover all the results of the
autonomous case. (2) We have proven that any time-dependent and
non-equilibrium solution of the nonautonomous master equation approaches its
unique steady equilibrium solution asymptotically. (3) The non-equilibrium
solution of the nonautonomous master equation is analysed numerically and
the different response behaviors of the system to the squeezing parameter is
exposed, and nonlinear effect is exhibited.

The works of dissipative system combined with the research of
control theory were extended in some publications
\cite{Jakob,Sten}, how to connect our approach with this
investigate is a interesting and challenge problem.

\section{Acknowledgment}

This work was supported in part by the National Natural Science Foundation
of China under grants No.10175029, 10375039, and 10004012, the Doctoral
Education Fund of the Education Ministry and Post-doctoral Science
Foundation, and the Nuclear Theory Research Fund of HIRFL of China.

\newpage

\begin{figure}[tbp]
\scalebox{1.0}{\includegraphics{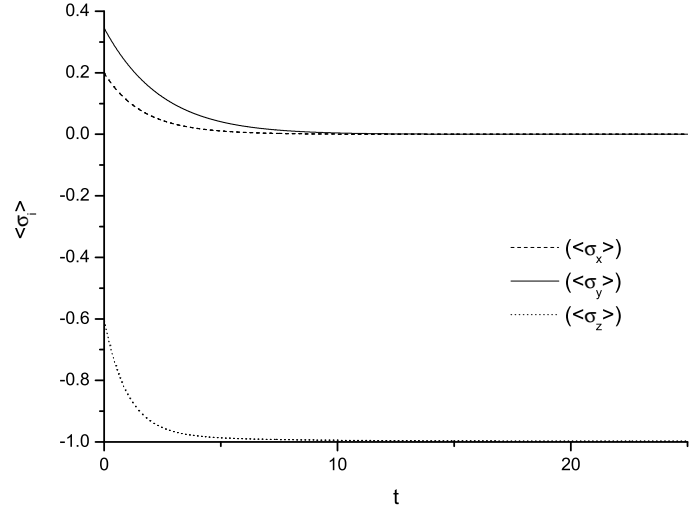}} \caption{The
time-dependence of $\langle \protect\sigma _{i}\rangle $ with
parameters $r=0.1e^{-0.1t}$, $\protect\mu =\protect\sqrt{0.2}e^{\protect\pi %
i/3}$, and $\protect\nu =\protect\sqrt{0.8}e^{2\protect\pi i}$.}
\label{Fig1}
\end{figure}

\begin{figure}[tbp]
\scalebox{1.0}{\includegraphics{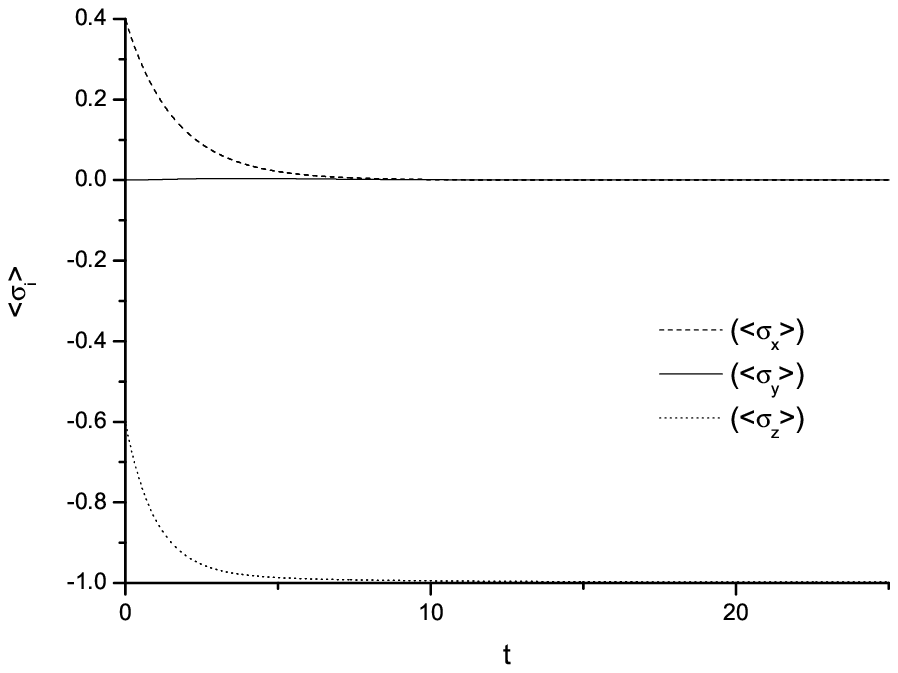}} \caption{The
time-dependence of $\langle \protect\sigma _{i}\rangle $ with
parameters $r=0.1e^{-0.1t}$, $\protect\mu =\protect\sqrt{0.2}$, and $\protect%
\nu =\protect\sqrt{0.8}$.}
\label{Fig2}
\end{figure}

\begin{figure}[tbp]
\scalebox{1.0}{\includegraphics{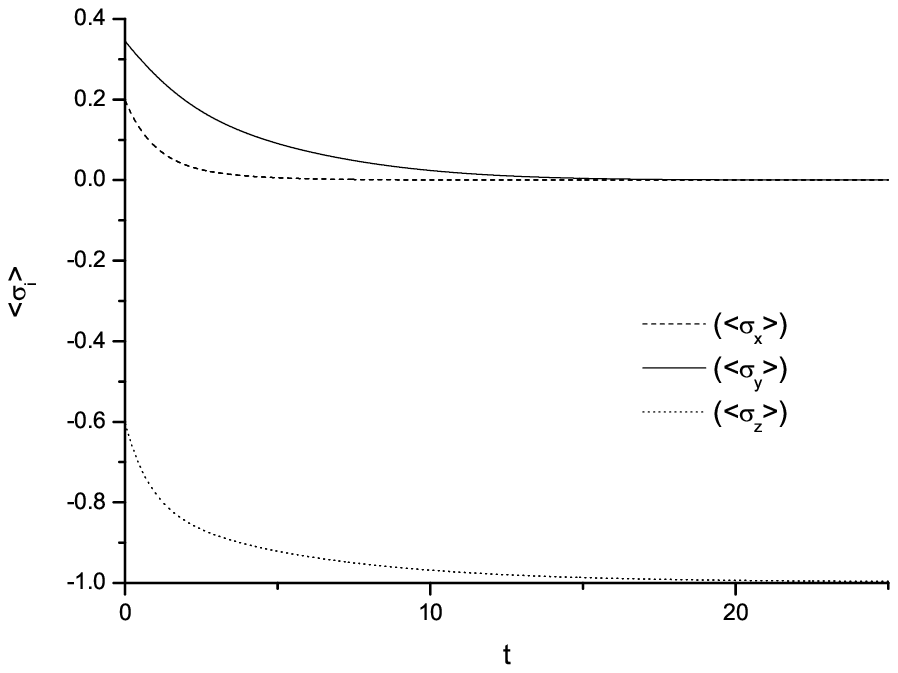}} \caption{The
time-dependence of $\langle \protect\sigma _{i}\rangle $ with
parameters $r=0.3e^{-0.1t}$, $\protect\mu =\protect\sqrt{0.2}e^{\protect\pi %
i/3}$, and $\protect\nu =\protect\sqrt{0.8}e^{2 \protect\pi i}$.}
\label{Fig3}
\end{figure}

\begin{figure}[tbp]
\scalebox{1.0}{\includegraphics{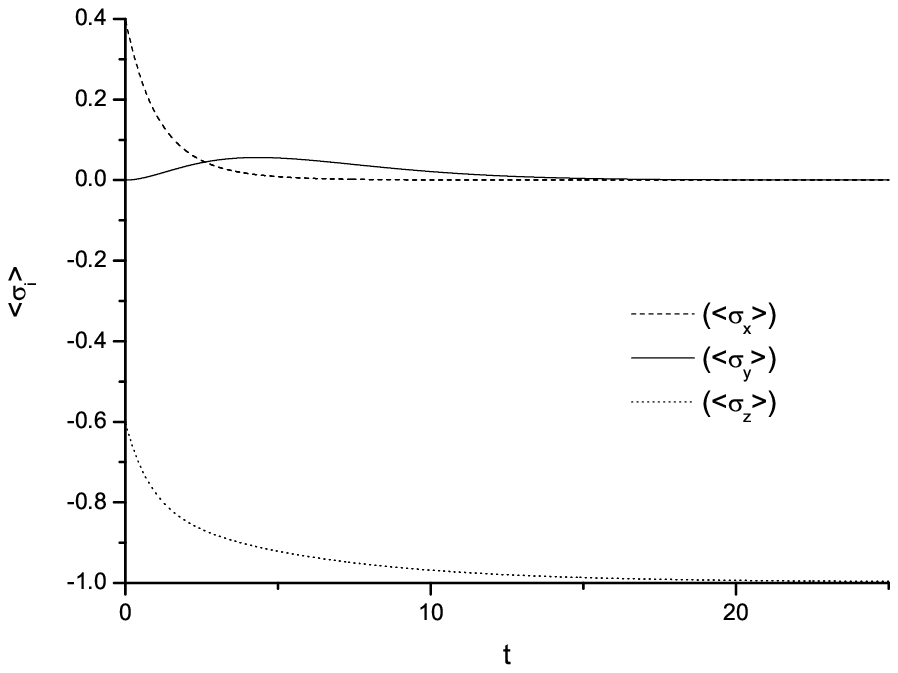}} \caption{The
time-dependence of $\langle \protect\sigma _{i}\rangle $ with
parameters $r=0.3e^{-0.1t}$, $\protect\mu =\protect\sqrt{0.2}$, and $\protect%
\nu =\protect\sqrt{0.8}$.}
\label{Fig4}
\end{figure}

\begin{figure}[tbp]
\scalebox{1.0}{\includegraphics{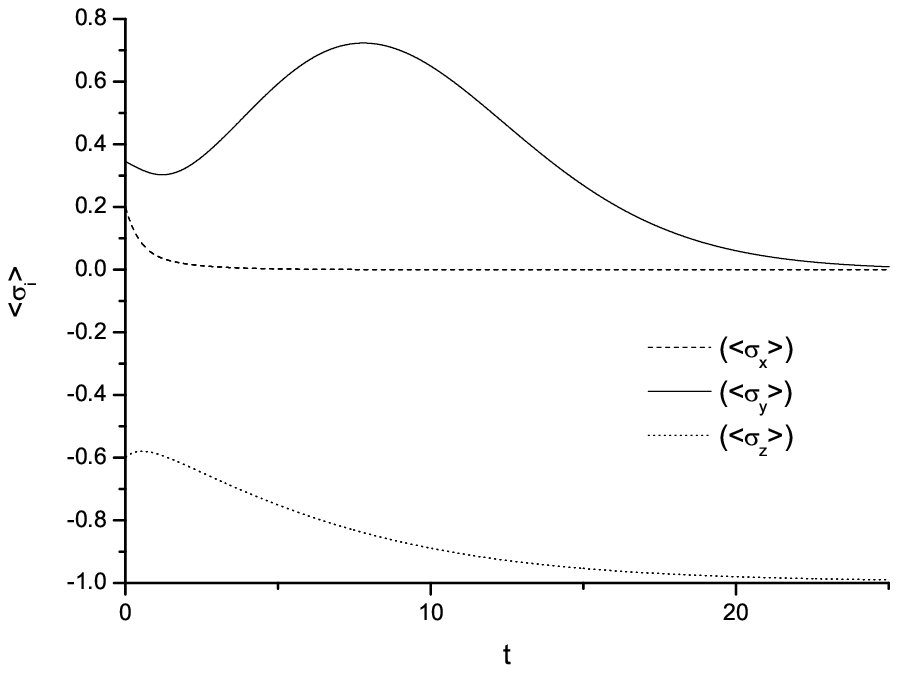}} \caption{The
time-dependence of $\langle \protect\sigma _{i}\rangle $ with
parameters $r=0.6e^{-0.1t}$, $\protect\mu =\protect\sqrt{0.2}e^{\protect\pi %
i/3}$, and $\protect\nu =\protect\sqrt{0.8}e^{2 \protect\pi i}$.}
\label{Fig5}
\end{figure}

\begin{figure}[tbp]
\scalebox{1.0}{\includegraphics{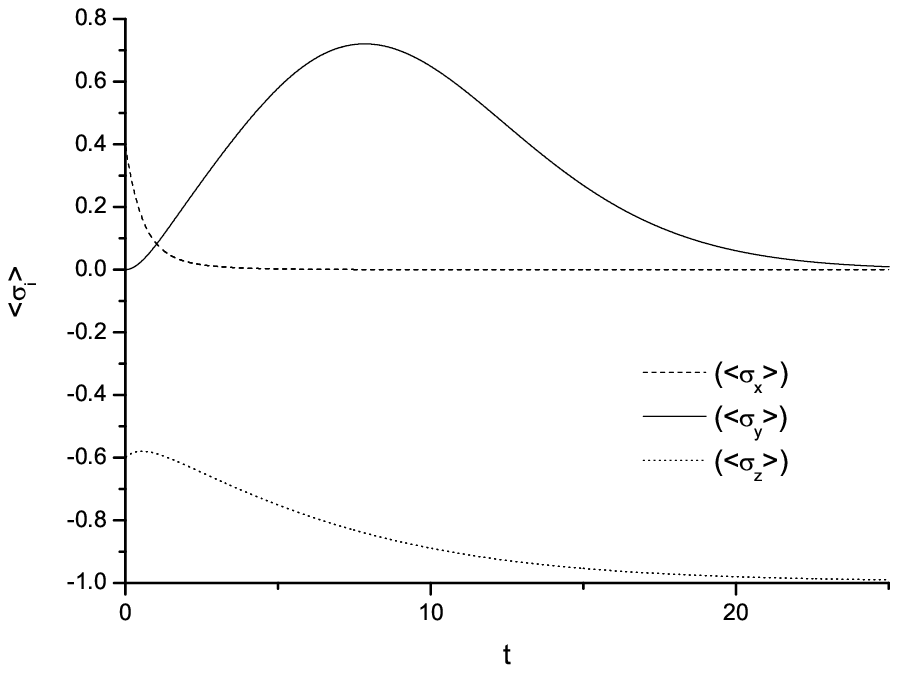}} \caption{The
time-dependence of $\langle \protect\sigma _{i}\rangle $ with
parameters $r=0.6e^{-0.1t}$, $\protect\mu =\protect\sqrt{0.2}$, and $\protect%
\nu =\protect\sqrt{0.8}$.}
\label{Fig6}
\end{figure}

\end{document}